\documentclass[11pt,a4paper]{article}

\usepackage{amsmath,amssymb,bm,bbm}
\usepackage{color}

\usepackage{graphicx}
\usepackage{hyperref}

\usepackage{tikz}

\newcommand{\R}{\mathbb{R}}

\newcommand{\cI}{\mathcal{I}}

\newcommand{\cZ}{\mathcal{Z}}

\newcommand{\U}{\mathrm{U}}

\newcommand{\Tr}{\mathrm{Tr}\,}

\newcommand{\Str}{\mathrm{Str}\,}
\newcommand{\Sdet}{\mathrm{Sdet}}
\newcommand{\id}{\mathbbm{1}}

\makeatletter

\@addtoreset{equation}{section}
\makeatother

\date{\today}

\begin{document}

\begin{titlepage}

\renewcommand{\thefootnote}{\fnsymbol{footnote}}

\begin{flushright}
 {\tt 
 IPHT-T14/118\\
 RIKEN-MP-94
 }
\\
\end{flushright}

\vskip9em

\begin{center}
 
 \begin{tikzpicture}
  
  \node (title) at (0,0)
  {\Large {\bf \hspace{-1.7em}
 Duality and integrability of supermatrix model with external source
 }};

 \end{tikzpicture}

 \vskip3em


 \setcounter{footnote}{1}
 {\sc Taro Kimura}\footnote{E-mail address: 
 \href{mailto:taro.kimura@cea.fr}
 {\tt taro.kimura@cea.fr}} 

 \vskip2em

{\it 
Institut de Physique Th\'eorique,
 CEA Saclay, 91191 Gif-sur-Yvette, France
 \\ \vspace{.5em}
Mathematical Physics Laboratory, RIKEN Nishina Center, 
 Saitama 351-0198, Japan 
}

 \vskip3em

\end{center}

 \vskip2em

\begin{abstract}
We study the Hermitian supermatrix model involving an external source.
We derive the determinantal formula for the supermatrix partition
 function, and also for the expectation value of the characteristic
 polynomial ratio, which yields the duality between the characteristic
 polynomial and the external source with an arbitrary matrix potential
 function.
We also show that the supermatrix integral satisfies the one
 and two dimensional Toda lattice equations as well as the ordinary
 matrix model.
\end{abstract}

\end{titlepage}

\tableofcontents

\setcounter{footnote}{0}


\section{Introduction}\label{sec:intro}

The matrix model has been playing a crucial role in quantum field theory
(QFT) for the last few decades, as a zero dimensional QFT model, or
rather a toy model for the infinite dimensional path integral.
Since it is just defined by finite dimensional integral of a matrix
itself, one can obtain an exact solution in various situations, which
provides a significant insight for the understanding of QFT.
This methodology is also refereed to as random matrix theory (RMT), and
is now applied to the extremely wide range of research
fields~\cite{Akemann:2011RMT}.

In QFT, in order to compute correlation functions, it is convenient to
introduce the generating function by adding the extra source term
\begin{equation}
 \cZ[J] = \int \mathcal{D}\phi \,
  e^{-\frac{1}{\hbar} S[\phi] + \int \! d^Dx \, J(x) \phi(x)}
  \, .
\end{equation}
The correlation function can be obtained by taking the functional
derivative with the source field
\begin{equation}
 \Big< \phi(x_1) \cdots \phi(x_k) \Big>
  = 
  \frac{\delta}{\delta J(x_1)} \cdots \frac{\delta}{\delta J(x_k)}
  \log \cZ[J] \Big|_{J=0} 
  \, .
\end{equation}
This generating function is defined in the sense of path integral, and
thus it is quite difficult to compute in general.
On the other hand, the matrix model having an external source, which is
just given by finite dimensional integral, plays a similar role to the
generating function in QFT,
\begin{equation}
 \cZ_N (A) = \int dX \, e^{-\frac{1}{\hbar} \Tr W(X) + \Tr XA}
  \, ,
  \label{MM_ext}
\end{equation}
where the integral is taken over a size $N$ matrix.
We can compute the correlation function by taking the derivative with
respect to the source matrix $A$, as well as the QFT generating
function.

A remarkable feature of the matrix model with the external source is the
duality with a characteristic polynomial, which was found
by~\cite{Brezin:2000CMP} especially for the Gaussian matrix model.
The claim of the duality is as follows: The $M$-point correlation function of
characteristic polynomials in the size $N$ matrix model with an external
source is equivalent to the $N$-point function with a size $M$ matrix,
under exchanging the arguments of the characteristic polynomial and the
external source.
This duality is now extended to generic
$\beta$-ensemble~\cite{Desrosiers:2008tp}, chiral
ensemble~\cite{Forrester:2013JPA}, and beyond the Gaussian
model~\cite{Kimura:2014mua}, and has various interesting interpretations
in terms of conformal field theory, string/M-theory, knot theory,
algebraic geometry, and so on.

Another interesting property of the external source matrix model is the
connection with the integrable system.
The matrix integral with an external source and also the characteristic
polynomial average satisfy the integrable equation, e.g., the Toda
lattice equation, and they can be interpreted as the corresponding
$\tau$-function.
See, for example,~\cite{Morozov:1994hh}.
Such integrability of the matrix model is quite useful to determine the
spectral density and also the correlation function.

In this paper we study a supersymmetric version of the matrix model
involving the external source based on a Hermitian supermatrix.
The supermatrix method is now well-known both in high energy physics and
condensed matter physics.
For example, one can avoid technical difficulty in dealing with the
normalization factor of the correlation function by applying the
supermatrix method, in a similar way to the replica method.
This property helps us to take the disorder average
in a random system~\cite{Efetov199610}.
In Sec.~\ref{sec:source} we will compute the $\U(N|N)$ symmetric
supermatrix partition function involving the external source.
As in the case of the ordinary Hermitian matrix model, we can utilize
the determinantal structure in order to perform the integral.
We will obtain a determinantal formula for the partition function, which
can be expressed as a size $N$ determinant of the $\U(1|1)$ partition
function as a determinantal kernel.

As in the case of the ordinary matrix model, there have been some proposals for
the duality of the supermatrix model.
For example, in particular for the Gaussian supermatrix model, one can
show an explicit duality between the external source and the
characteristic polynomial~\cite{Desrosiers:2009pz}.
In Sec.~\ref{sec:ch_poly} we will exhibit that this duality generally
holds for the supermatrix with an arbitrary potential function, and
is just interpreted as Fourier transformation.
In addition, it will be pointed out that the relation between the
characteristic polynomial ratio with the ordinary matrix
model~\cite{Fyodorov:2002jw} and supermatrix
models~\cite{Zirnbauer:1996zz,Brezin:2003JPA,Kieburg:2009jd} is naturally
explained as some specialization of the characteristic polynomial
expectation value in the supermatrix.

In Sec.~\ref{sec:integrable} we will also show the integrable equations
for the supermatrix integral with the external source and the
characteristic polynomial, based on the determinantal formula derived in
this article.
Since we have more parameters for the partition function rather than the
ordinary matrix model, we can obtain several integrable equations
corresponding to variables parametrizing the external source and the
characteristic polynomial.

\section{Supermatrix model with external source}\label{sec:source}

Let us introduce a Hermitian supermatrix model with a potential
function $W(x)$, involving the external source matrix $C$, as a natural
generalization of the Hermitian matrix model~(\ref{MM_ext}),
\begin{equation}
 \cZ_{N,M} (\{a_i\}_{i=1}^N, \{b_j\}_{j=1}^M)
  =
  \int dZ \, e^{-\frac{1}{\hbar} \Str W(Z) + \Str Z C}
  \, ,
  \label{SMM01}
\end{equation}
where $\hbar$ is the coupling constant.
The Hermitian supermatrices $Z$ and $C$ are given by
\begin{equation}
 Z =
  \left(
   \begin{array}{cc}
    X & \xi \\ \xi^\dag & Y
   \end{array}
  \right)
  \, , \qquad
 C =
 \left(
  \begin{array}{cc}
   A & \eta \\ \eta^\dag & B
  \end{array}
 \right) 
 \, ,
\end{equation}
where $X$, $A$ are $N \times N$, and $Y$, $B$ are $M \times M$
(bosonic) Hermitian matrices.
$\xi$ and $\eta$ are $N \times M$ fermionic matrices, whose elements are
given by Gra{\ss}mannian variables.
We can assume $C$ is a diagonal matrix without loss of generality, and
in this paper we also assume $N \ge M$.

The matrix measure of (\ref{SMM01}) is invariant under the supergroup
transformation, $dZ = d(U Z U^{-1})$ with $U \in \U(N|M)$, and is
normalized by the corresponding volume factor $\operatorname{Vol.} \U(N|M)$.
As well as the ordinary Hermitian matrix integral, one can decompose
this measure into the diagonal and angular parts
\begin{equation}
 dZ = \Delta_{N,M}(x;y)^2
  d^N x \, d^N y \, dU 
  \, ,
\end{equation}
and the corresponding Jacobian $\Delta_{N,M}(x;y)$ is now given by the
generalized Cauchy determinant~\cite{Basor:1994MN}
\begin{align}
 \Delta_{N,M}(x;y)
 &\equiv
 \Delta_N(x) \Delta_M(y)
  \prod_{i=1}^N \prod_{j=1}^M (x_i - y_j)^{-1}
 \nonumber \\
 & =
  \det
  \left(
   \begin{array}{c}
    \displaystyle x_i^{k-1} \\ 
    \displaystyle (x_i-y_j)^{-1}
   \end{array}
  \right)
 \quad
 \mbox{with}
 \quad
 \begin{cases}
  i = 1, \cdots, N \\
  j = 1, \cdots, M \\
  k = 1, \cdots, N-M
 \end{cases}
 \, ,
 \label{Cauchy_det}
\end{align}
where $\Delta_N(x)$ is the Vandermonde determinant
\begin{equation}
 \Delta_N(x) 
  = \det_{1 \le i, j \le N} x_i^{j-1}
  = \prod_{i<j}^N (x_i - x_j)
  \, .
\end{equation}
The fact that the Jacobian factor can be written as a determinant will
play a crucial role in the following discussion.

To perform the angular part integral of (\ref{SMM01}), we now apply
the supergroup version of the Harish-Chandra--Itzykson--Zuber (HCIZ)
formula, which is written in terms of the Cauchy
determinant~\cite{Guhr:1991JMP,Alfaro:1994ca,Guhr:1996CMP},
\begin{align}
 \cI_{N,M}(Z,C)
 & \equiv
 \int_{\U(N|M)} dU \, e^{\Str Z U C U^{-1}}
 \nonumber \\
 &
  =
  \frac{1}{\Delta_{N,M}(x;y) \Delta_{N,M}(a;b)}
  \det_{1 \le i, j \le N} e^{x_i a_j}
  \det_{1 \le i, j \le M} e^{- y_i b_j}  
  \, .
 \label{HCIZ_U(N|M)}
\end{align}
This is a quite natural extension of the original formula for the $\U(N)$
group, which is given by replacing the Cauchy determinant with the
Vandermonde determinant~\cite{HC:1957,Itzykson:1979fi},
\begin{eqnarray}
 \cI_N (X,A) 
  & \equiv &
  \int_{\U(N)} dU \, e^{\Tr X U A U^{-1}}
  \nonumber \\
 & = &
  \frac{1}{\Delta_N(x) \Delta_N(a)}
  \det_{1 \le i, j \le N} e^{x_i a_j} \, .
  \label{HCIZ_U(N))}
\end{eqnarray}
Applying the formula (\ref{HCIZ_U(N|M)}), we obtain the expression only
in terms of eigenvalues
\begin{align}
 \cZ_{N,M}
  & =
  \frac{1}{\Delta_{N,M}(a;b)}
  \int
  \prod_{i=1}^N dx_i \,
  e^{-\frac{1}{\hbar} W(x_i) + x_i a_i}
  \prod_{j=1}^M dy_j \,
  e^{\frac{1}{\hbar} W(y_j) - y_j b_j} \,
  \Delta_{N,M}(x;y)
 \, .
\end{align}
Let us comment on convergency of this integral.
Naively speaking, it is not convergent integral in general.
We can regard this as a formal integral,
but one possible way to avoid this divergence is to modify the matrix
potential by inserting a constant supermatrix $I_{N,M} = \mathrm{diag}(\id_N,
i \id_M) \in \U(N|M)$, i.e., $
\Str Z^2 = \Tr X^2 - \Tr Y^2 \to \Str I_{N,M} Z^2 = \Tr X^2 + \Tr
Y^2$~\cite{Desrosiers:2008tp}.
However, this treatment does not work for the generic potential function
$W(x)$.
Another possibility is considering the coupling constant as an imaginary
number, $\hbar \in i \R$.
In this case, although the integrand becomes an oscillating function,
one can provide interpretation as the Fresnel integral.
For example, this interpretation works well for the Chern--Simons matrix
model~\cite{Marino:2002fk} and its supermatrix generalization, which is
called the ABJM matrix
model~\cite{Kapustin:2009kz,Drukker:2009hy,Marino:2009jd}, although the
corresponding coupling constant is pure imaginary, e.g., $\hbar = 2 \pi
i/(k+N)$, or $\hbar = 2 \pi i / k$.

Then, due to the Cauchy determinantal formula (\ref{Cauchy_det}), it can
be written as a size $N$ determinant, consisting of $N \times M$ and $N
\times (N-M)$ blocks, 
\begin{equation}
 \cZ_{N,M} (\{a_i\}_{i=1}^N,\{b_j\}_{j=1}^M)
  =
 \frac{1}{\Delta_{N,M}(a;b)}
 \det 
 \left(
 \begin{array}{c}
  Q_{k-1}(a_i) \\ R(a_i;b_j)
 \end{array}
 \right)
 \, ,
 \label{SMM02}
\end{equation}
where the indices run as $i = 1, \cdots, N$, $j = 1, \cdots, M$, $k = 1,
\cdots, N-M$, and we have introduced the following functions,
\begin{align}
 P_{i-1}(x)
 & =
 x^{i-1} 
 \, , \quad
 \tilde{R}(x;y) = \frac{1}{x-y}
 \, , \label{P-func} \\
 Q_{i-1}(a) 
 & = 
 \int dx \, P_{i-1}(x) \, e^{-\frac{1}{\hbar} W(x) + xa} \, ,
 \label{Q-func}
 \\
 R(a;b)
 & =
 \int dx dy \,
 \tilde{R}(x;y) \, e^{-\frac{1}{\hbar} W(x) + \frac{1}{\hbar} W(y) + xa - yb} 
 \, .
 \label{R-func}
\end{align}
Note that the function (\ref{Q-func}) obeys 
\begin{equation}
 Q_{i}(a) = \frac{d}{d a} Q_{i-1}(a) \, ,
\end{equation}
and $Q_{i=0}(a)$ is seen as a generalized Airy function
\begin{equation}
 Q_{i=0}(a) = \int dx \, e^{-\frac{1}{\hbar} W(x) + x a} \, .
\end{equation}
As we will see later, these functions are just seen as (double) Fourier
or rather Laplace transform of $\displaystyle P_{i-1}(x)
e^{-\frac{1}{\hbar}W(x)}$ and $\displaystyle \tilde{R}(x;y)
e^{-\frac{1}{\hbar} W(x) + \frac{1}{\hbar} W(y)}$, respectively.%
\footnote{
If we modify the source term in the supermatrix integral
\begin{equation}
 \cZ_{N,M} (\{a_i\}_{i=1}^N, \{b_j\}_{j=1}^M)
  =
  \int dZ \, e^{-\frac{1}{\hbar} \Str W(Z) + i \, \Str Z C}
  \, ,
\end{equation}
they just become Fourier transforms:
\begin{equation}
 Q_{i-1}(a) = \int dx \, P_{i-1}(x) \,
  e^{-\frac{1}{\hbar} W(x) + i x a}
  \, , \quad
 R(a;b)
 =
 \int dx dy \,
 \tilde{R}(x;y) \,
 e^{-\frac{1}{\hbar} W(x) + \frac{1}{\hbar} W(y) + i xa - i yb} \, .
\end{equation}
}

Let us comment on some specialization of the formula (\ref{SMM02}).
If we take the limit $M=0$, the determinantal expression (\ref{SMM02})
is reduced to the the well-known formula for the ordinary Hermitian
matrix model with the external source
\begin{align}
 \cZ_{N} (\{a_i\}_{i=1}^N)
 & =
  \int \! dX \, e^{-\frac{1}{\hbar} \Tr W(X) + \Tr XA}
 \nonumber \\
 & =
  \frac{1}{\Delta_N(a)} \det_{1 \le i, j \le N} Q_{j-1}(a_i)
 \, .
\end{align}
For example, see \cite{Morozov:1994hh} for details.
Then, in the case with $M=N$, the partition function becomes
\begin{align}
 \cZ_{N,N} (\{a_i\}_{i=1}^N,\{b_j\}_{j=1}^N)
 & =
 \frac{1}{\Delta_{N,N}(a;b)}
 \det_{1 \le i, j \le N} 
 R(a_i;b_j)
 \, .
\end{align}
This shows that the formula for $\U(N|N)$ theory is reduced to a size $N$
determinant of the ``dual'' kernel function $R(a;b)$, which is given by
the $\U(1|1)$ partition function
\begin{equation}
 R(a;b) = \frac{\cZ_{1,1} (a,b)}{a-b}  \, .
\end{equation}
This factorization property is called Giambelli
compatibility~\cite{Borodin:2006AAM}, and follows the Fay
identity~\cite{Fay:1973}.

\section{Characteristic polynomial average and duality}\label{sec:ch_poly}

In this section we consider expectation values of the
characteristic polynomial ratio with the supermatrix model in the
presence of an external source
\begin{align}
 & 
 \Psi_{N,M;\,p,q} (\{a_i\}_{i=1}^N, \{b_j\}_{j=1}^M; 
 \{\lambda_\alpha\}_{\alpha=1}^p, \{\mu_\beta\}_{\beta=1}^q)
 \nonumber \\
 & \hspace{5em}
 =
  \int \! dZ \, e^{-\frac{1}{\hbar} \Str W(Z) + \Str Z C}
  \prod_{\alpha=1}^p \Sdet \left( \lambda_\alpha - Z \right)
  \prod_{\beta=1}^q \Sdet \left( \mu_\beta - Z \right)^{-1}
 \, ,
 \label{EV_ratio01}
\end{align}
and then show the duality between the $(p|q)$-point function with
$\U(N|M)$ theory and $(N|M)$-point function with $\U(p|q)$ theory.
The characteristic polynomial average (\ref{EV_ratio01}) includes
several useful situations:
\begin{itemize}
 \item $M=q=0$: Characteristic polynomial with the Hermitian matrix model
       \begin{equation}
	\Psi_{N;\,p} (\{a_i\}_{i=1}^N; \{\lambda_\alpha\}_{\alpha=1}^p)
	 = 
	 \int dX \,
	 e^{-\frac{1}{\hbar}\Tr W(X) + \Tr XA}
	 \prod_{\alpha=1}^p \det (\lambda_\alpha - X)
	 \, .
	 \label{M=q=0}
       \end{equation}
 \item $M=p=0$: Characteristic polynomial inverse 
       with the Hermitian matrix model
       \begin{equation}
	\Psi_{N;\,q} (\{a_i\}_{i=1}^N; \{\mu_\beta\}_{\beta=1}^q)
	 = 
	 \int dX \,
	 e^{-\frac{1}{\hbar}\Tr W(X) + \Tr XA}
	 \prod_{\beta=1}^q \det (\mu_\beta - X)^{-1}
	 \, .
	 \label{M=p=0}
       \end{equation}
 \item $M=0$: Characteristic polynomial ratio with the Hermitian matrix model
       \begin{align}
	&
	\Psi_{N;\,p, q} 
	 (\{a_i\}_{i=1}^N; 
	 \{\lambda_\alpha\}_{\alpha=1}^p, \{\mu_\beta\}_{\beta=1}^q)
	\nonumber \\
	& \hspace{5em}
	 = 
	 \int dX \,
	 e^{-\frac{1}{\hbar}\Tr W(X) + \Tr XA}
	 \prod_{\alpha=1}^p \det (\lambda_\alpha - X)
	 \prod_{\beta=1}^q \det (\mu_\beta - X)^{-1}
	 \, .
	\label{M=0}
       \end{align}
\end{itemize}
Therefore the expression (\ref{EV_ratio01}) provides a {\em master formula}
for the characteristic polynomial average in various matrix models.
For example, the duality in the case with $M=0$ (\ref{M=0}) claims that the
characteristic polynomial ratio with the ordinary Hermitian matrix model
is dual to another supermatrix model, as shown in~\cite{Brezin:2003JPA}.

As discussed in Sec.~\ref{sec:source}, the angular part of the
integral is performed using the HCIZ formula~(\ref{HCIZ_U(N|M)}), and then we
obtain the expression only in terms of eigenvalues,
\begin{align}
 \Psi_{N,M;\,p,q}
 & =
 \frac{1}{{\Delta}_{N,M}(a;b)}
 \int 
 \prod_{i=1}^N dx_i \, e^{-\frac{1}{\hbar} W(x_i) + x_i a_i}
 \prod_{j=1}^M dy_j \, e^{\frac{1}{\hbar} W(y_j) - y_j b_j}
 \, \Delta_{N,M}(x;y)
 \nonumber \\
 &
 \hspace{10em} \times
 \prod_{\alpha=1}^p
 \left(
 \frac{\prod_{i=1}^N (\lambda_\alpha - x_i)}
      {\prod_{j=1}^M (\lambda_\alpha - y_j)}
 \right)
 \prod_{\beta=1}^q
 \left(
 \frac{\prod_{j=1}^M (\mu_\beta - y_j)}
      {\prod_{i=1}^N (\mu_\beta - x_i)}     
 \right)
  \, .
\end{align}
It is now convenient to apply the following identity to this expression,
\begin{equation}
 {\Delta}_{N+p,M+q}(x,\lambda;y,\mu)
  = 
  {\Delta}_{N,M} (x;y)
  {\Delta}_{p,q} (\lambda;\mu)
 \prod_{\alpha=1}^p
 \left(
 \frac{\prod_{i=1}^N (\lambda_\alpha - x_i)}
      {\prod_{j=1}^M (\lambda_\alpha - y_j)}
 \right)
 \prod_{\beta=1}^q
 \left(
 \frac{\prod_{j=1}^M (\mu_\beta - y_j)}
      {\prod_{i=1}^N (\mu_\beta - x_i)}     
 \right)
 \, .
\end{equation}
If we assume $N+p > M+q$, the LHS of this equation is a large Cauchy
determinant of the size $N+p$, with several block structures,
\begin{align}
 \Delta_{N+q, M+q}(a,\lambda; b,\mu)
 & =
 \det 
 \left(
  \begin{array}{cc}
   x_i^{k-1} & \lambda_\alpha^{k-1} \\
   \left(x_i - y_j\right)^{-1} & \left(\lambda_\alpha - y_j\right)^{-1} \\
   \left(x_i - \mu_\beta\right)^{-1} 
    & \left(\lambda_\alpha - \mu_\beta\right)^{-1}
  \end{array}
 \right)
 \\
 & = (-1)^{Np+Mq}
 \det 
 \left(
  \begin{array}{cc}
   \lambda_\alpha^{k-1} & x_i^{k-1} \\
   \left(\lambda_\alpha - \mu_\beta\right)^{-1} 
    & \left(x_i - \mu_\beta\right)^{-1} \\
   \left(\lambda_\alpha - y_j\right)^{-1} & \left(x_i - y_j\right)^{-1} \\
  \end{array}
 \right)
 \, ,
\end{align}
where $i = 1, \cdots, N$, $j = 1, \cdots, M$, 
$\alpha = 1, \cdots, p$, $\beta = 1, \cdots, q$, and $k = 1, \cdots, N+p-M-q$.
Due to this expression, we obtain the determinantal formula for the
characteristic polynomial ratio expectation value
\begin{align}
 &
 \Psi_{N,M;\,p,q} 
 ( \{a_i\}_{i=1}^N, \{b_j\}_{j=1}^M; 
   \{\lambda_\alpha\}_{\alpha=1}^p, \{\mu_\beta\}_{\beta=1}^q )
 \nonumber \\
 & =
 \frac{1}{{\Delta}_{N,M}(a;b) {\Delta}_{p,q}(\lambda;\mu)}
 \int 
 \prod_{i=1}^N dx_i \, e^{-\frac{1}{\hbar} W(x_i) + x_i a_i}
 \prod_{j=1}^M dy_j \, e^{\frac{1}{\hbar} W(y_j) - y_j b_j}
 \, \Delta_{N+p,M+q}(x,\lambda; y,\mu)
 \nonumber \\
 & =
 \frac{1}{{\Delta}_{N,M}(a;b) {\Delta}_{p,q}(\lambda;\mu)}
 \det
 \left(
 \begin{array}{cc}
  Q_{k-1} (a_i) & P_{k-1}(\lambda_\alpha) \\
  R(a_i;b_j) & S_{\rm R} (\lambda_\alpha;b_j) \\
  S_{\rm L} (a_i;\mu_\beta) & \tilde{R}(\lambda_\alpha;\mu_\beta) \\
 \end{array}
 \right)
 \, ,
 \label{SChPoly_det01}
\end{align}
where we have introduced the following auxiliary functions in addition
to (\ref{P-func}), (\ref{Q-func}) and (\ref{R-func}),
\begin{equation}
 S_{\rm L} (a;\mu)
 =
  \int \! dx \,
  \frac{1}{x-\mu} \, e^{-\frac{1}{\hbar} W(x) + xa}
  \, , \quad
 S_{\rm R} (\lambda;b)
 =
  \int \! dy \,
  \frac{1}{\lambda - y} \, e^{\frac{1}{\hbar} W(y) - y b}
 \label{R_tilde}\, .
\end{equation}
The formula (\ref{SChPoly_det01}) shows a duality between the external
source and the characteristic polynomials in the supermatrix model with
an arbitrary potential function $W(x)$, since it can be also written as
follows
\begin{align}
 &
 \Psi_{N,M;\,p,q} 
 ( \{a_i\}_{i=1}^N, \{b_j\}_{j=1}^M; 
   \{\lambda_\alpha\}_{\alpha=1}^p, \{\mu_\beta\}_{\beta=1}^q )
 \nonumber \\
  & \hspace{3em} 
 =
 \frac{(-1)^{Np+Mq}}{{\Delta}_{N,M}(a;b) {\Delta}_{p,q}(\lambda;\mu)}
 \det
 \left(
 \begin{array}{cc}
  P_{k-1}(\lambda_\alpha) & Q_{k-1} (a_i) \\
  \tilde{R}(\lambda_\alpha;\mu_\beta) & S_{\rm L} (a_i;\mu_\beta) \\
  S_{\rm R} (\lambda_\alpha;b_j) & R(a_i;b_j) \\
 \end{array}
 \right)
 \, .
\end{align}
This duality was originally shown in the
Gaussian model with the harmonic potential $W(x) = \frac{1}{2}
x^2$~\cite{Desrosiers:2009pz}.
Under this duality, we see the correspondence between the auxiliary
functions:
\begin{equation}
 P_k(\lambda) 
  \ \longleftrightarrow \
 Q_k(a)
 \, , \quad
 S_{\rm L}(a;\mu)
  \ \longleftrightarrow \
 S_{\rm R}(\lambda;b)
 \, , \quad
 R(a;b)
  \ \longleftrightarrow \
 \tilde{R}(\lambda;\mu)
 \, ,
\end{equation}
and then the $(p|q)$-point function for $\U(N|M)$
supermatrix model with the potential $W(x)$ is converted into the
$(N|M)$-point function for $\U(p|q)$ supermatrix model with another
potential $\tilde{W}(x)$, which is obtained through Fourier transformation:
\begin{align}
 &
 \Psi_{N,M;\,p,q}
 ( \{a_i\}_{i=1}^N, \{b_j\}_{j=1}^M; 
   \{\lambda_\alpha\}_{\alpha=1}^p, \{\mu_\beta\}_{\beta=1}^q )
 \nonumber \\
 & \hspace{3em} \stackrel{\mbox{\scriptsize F.T.}}{=}
 {\Psi}_{p,q;N,M}
 ( \{\lambda_\alpha\}_{\alpha=1}^p, \{\mu_\beta\}_{\beta=1}^q;
   \{a_i\}_{i=1}^N, \{b_j\}_{j=1}^M )
   \, .
 \label{BH-duality}
\end{align}
In this sense, it is easier to see this duality especially for the
Gaussian matrix model, because the Gaussian function is self-dual with
respect to Fourier transformation.
We remark that this kind of duality can be also found in the Chern--Simons
matrix model and the ABJM matrix model, which are obtained by replacing
the Vandermonde determinant with the exponentiated one,
\begin{equation}
 \Delta(x) =
 \prod_{i<j}^N (x_i - x_j)
  \ \longrightarrow \
  \prod_{i<j}^N \left( 2 \sinh \frac{x_i - x_j}{2} \right)
  \, .
\end{equation}
In this case insertion of an external source implies the 
expectation value of the Wilson loop operator, which is characterized by
Schur function~\cite{Eynard:2014rba}.

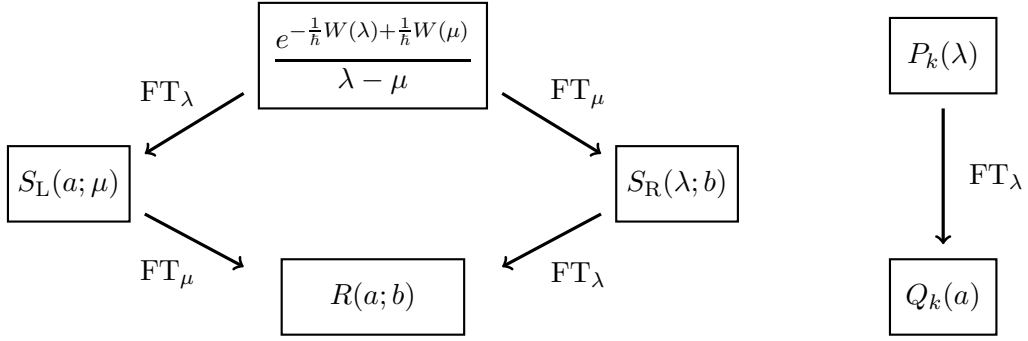
\begin{figure}[t]
 \begin{center}
   \begin{tikzpicture}
  
  \node (0) at (0,0) 
    {$\displaystyle 
    \frac{ e^{ -\frac{1}{\hbar} W(\lambda) + \frac{1}{\hbar} W(\mu)} }
         {\lambda - \mu}$};

  \node (L) at (-4,-1.7)
    {$\displaystyle S_{\rm L}(a;\mu)$};

  \node (R) at (4,-1.7)
    {$\displaystyle S_{\rm R}(\lambda;b)$};

  \node (LR) at (0,-3.2)
    {$\displaystyle R(a;b)$};

  \node (P) at (7.5,0)
    {$\displaystyle P_k(\lambda)$};

  \node (Q) at (7.5,-3.2)
    {$\displaystyle Q_k(a)$};

  \draw [thick] (-1.5,-.7) rectangle (1.5,.7);
  \draw [thick] (-4.8,-2.2) rectangle (-3.2,-1.2);
  \draw [thick] (4.8,-2.2) rectangle (3.2,-1.2);
  \draw [thick] (-1.2,-3.7) rectangle (1.2,-2.7);

  \draw [thick] (6.8,-0.5) rectangle (8.2,0.5);
  \draw [thick] (6.8,-3.7) rectangle (8.2,-2.7);

  \draw [very thick,->] (-1.7,-0.5) -- (-3,-1.3);
  \draw [very thick,->] (1.7,-0.5) -- (3,-1.3);
  \draw [very thick,->] (-3,-2.1) -- (-1.7,-2.8);
  \draw [very thick,->] (3,-2.1) -- (1.7,-2.8);

  \draw [very thick,->] (7.5,-0.7) -- (7.5,-2.5);

  \node (FT_L1) at (-2.7,-.5) {FT$_\lambda$};
  \node (FT_L2) at (-2.7,-2.9) {FT$_\mu$};

  \node (FT_R1) at (2.7,-.5) {FT$_\mu$};
  \node (FT_R2) at (2.7,-2.9) {FT$_\lambda$};

  \node (FT) at (8.2,-1.6) {FT$_\lambda$};

\end{tikzpicture}
 \end{center}
 \caption{{\em Fourier transform web}: the relation between the
 auxiliary functions. FT$_x$ means Fourier transformation with respect to
 $x$ variable. 
 The duality (\ref{BH-duality}) is seen as the double Fourier
 transformation with $\lambda$ and $\mu$ variables.
 }
 \label{FT_flow}
\end{figure}

To see this duality more explicitly, we now introduce another
correlation function by multiplying the weight functions
\begin{align}
 &
 \Phi_{N,M;\,p,q} 
 ( \{a_i\}_{i=1}^N, \{b_j\}_{j=1}^M; 
   \{\lambda_\alpha\}_{\alpha=1}^p, \{\mu_\beta\}_{\beta=1}^q )
 \nonumber \\
 & \hspace{3em}
   =
 \Psi_{N,M;\,p,q} 
 ( \{a_i\}_{i=1}^N, \{b_j\}_{j=1}^M; 
   \{\lambda_\alpha\}_{\alpha=1}^p, \{\mu_\beta\}_{\beta=1}^q )
   \prod_{\alpha=1}^p e^{-\frac{1}{\hbar} W(\lambda_\alpha)}
   \prod_{\beta=1}^q e^{\frac{1}{\hbar} W(\mu_\beta)}
 \, .
\end{align}
This can be also written in the determinantal form
(\ref{SChPoly_det01}), but by replacing the auxiliary functions
as follows,
\begin{align}
 &
 P_k(\lambda)
 \ \to \
 \lambda^k \, e^{\mp \frac{1}{\hbar} W(\lambda)}
 \, , \quad
 Q_k(a)
  \ \to \
 \int dx \, x^k \, e^{\mp \frac{1}{\hbar} W(x) \pm xa}
 \, , \label{PQ-func} \\
 &
 R(a;b)
  \ \to \
 \int dx dy \,
 \frac{1}{x-y} \,
 e^{-\frac{1}{\hbar} W(x) + \frac{1}{\hbar} W(y) + xa - yb}
 \, , \quad
 \tilde{R}(\lambda;\mu)
 \ \to \
 \frac{1}{\lambda - \mu} \, 
 e^{-\frac{1}{\hbar} W(\lambda) + \frac{1}{\hbar} W(\mu)} 
 \, , \\
 &
 S_{\rm L}(a;\mu)
 \ \to \
 \int dx \,
 \frac{1}{x-\mu} \,
 e^{-\frac{1}{\hbar} W(x) + \frac{1}{\hbar} W(\mu) + xa} 
 \, , \quad
 S_{\rm R}(\lambda;b)
 \ \to \
 \int dy \,
 \frac{1}{\lambda - y} \,
 e^{-\frac{1}{\hbar} W(\lambda) + \frac{1}{\hbar} W(y) - yb} 
 \, .
\end{align}
The sign factor in (\ref{PQ-func}) depends on whether $M + p > N + q$ or
$M + p < N + q$.
In this case it is easier to see the relation between auxiliary
functions, which is summarized in Fig.~\ref{FT_flow}.

\section{Integrable equations}\label{sec:integrable}

In the previous sections, we have shown that the partition functions of
the Hermitian supermatrix models have the determinantal formula, and
such a determinant structure plays an important role in the relation to
the integrable systems~\cite{Morozov:1994hh}:
it behaves as a $\tau$-function for the corresponding
integrable equation.
In this section we study the connection between the supermatrix
partition function and the integrable system.

\subsection*{Supermatrix model with external source}

We now show that the supermatrix integral involving the external source
(\ref{SMM02}) can be interpreted as the $\tau$-function for the Toda
lattice equations.
In this paper we especially derive the integrable equation, obtained by
taking all the external source parameters to the same value.

We first split external source parameters into the center of mass and
relative parts
\begin{equation}
 a_i = \delta a_i + a \, , \qquad
 b_j = \delta b_j + b \, , \qquad
 \begin{cases}
  i = 1, \cdots, N \\ j = 1, \cdots, M
 \end{cases}
 \, .
\end{equation}
With this separation the determinant (\ref{Cauchy_det}) is given by
\begin{equation}
 \Delta_{N,M}(a;b)
  =
  \Delta_N (\delta a) \Delta_M (\delta b)
  \prod_{i=1}^N \prod_{j=1}^M
  (\delta a_i - \delta b_j + a - b)^{-1}
  \, .
\end{equation}
Let us then consider the Taylor expansion around the center of mass for
the functions (\ref{Q-func}) and (\ref{R-func}),
\begin{align}
 Q_{k-1}(a_i)
 & =
  \sum_{l=1}^\infty \frac{(\delta a_i)^{l-1}}{(l-1)!} \, Q_{k+l-2}(a)
 \, , \\
 R(a_i;b_j)
 & =
 \sum_{l,m=1}^\infty 
 \frac{(\delta a_i)^{l-1}}{(l-1)!} 
 \frac{(\delta b_j)^{m-1}}{(m-1)!} \,
 R^{(l-1,m-1)}(a;b)
 \, ,
\end{align}
with
\begin{equation}
 R^{(i,j)}(a;b) 
  = 
  \frac{\partial^{i}}{\partial a^i}
  \frac{\partial^j}{\partial b^j} R(a;b)
  \, .
\end{equation}
By taking the limit $\delta a_i = \delta b_j = 0$ for $\forall i,
j$, the determinant in (\ref{SMM02}) is written
\begin{align}
 \det
 \left(
 \begin{array}{c}
  Q_{k-1} (a_i) \\ R(a_i; b_j)
 \end{array}
 \right)
 =
 \det_{1 \le i, j \le N} \frac{\left(\delta a_i\right)^{j-1}}{(j-1)!}
 \cdot
 \det_{1 \le i, j \le M} \frac{\left(\delta b_i\right)^{j-1}}{(j-1)!}
 \cdot
 \det
 \left(
 \begin{array}{c}
  Q_{i+k-2} (a) \\ R^{(i-1,j-1)} (a; b)
 \end{array}
 \right)
 \, ,
\end{align}
where the range of indices is given by $i=1, \cdots, N$, $j = 1,
\cdots, M$, and $k = 1, \cdots, N - M$ in the determinants of the LHS
and the third one in the RHS.
Therefore, after some cancellation, we obtain the following formula 
\begin{align}
 &
 \cZ_{N,M} (\{a_i=a\}_{i=1}^N,\{b_j=b\}_{j=1}^M)
  = 
  c_{N,M} \, \tilde{\cZ}_{N,M}(a,b) \, , \\
 & 
 \tilde{\cZ}_{N,M}(a,b)
 =
  \det \left(
	\begin{array}{c}
	 Q_{i+k-2}(a) \\ R^{(i-1,j-1)}(a;b)
	\end{array}
       \right)
  \, ,
 \label{SMM03}
\end{align}
with the constant
\begin{equation}
 c_{N,M} = 
  (a-b)^{NM}
  G(N+1)^{-1} G(M+1)^{-1}
  \, ,
\end{equation}
where $\displaystyle G(n) = \prod_{i=0}^{n-2} i! = \prod_{i=1}^{n-1}
\Gamma(i)$ is the Barnes G-function.
We remark that this factor corresponds to the volume element for $\U(N)$ and
$\U(M)$ groups.
This shows that the supermatrix integral (\ref{SMM01}) is finally written as a
kind of Wronskian in the equal parameter limit, $\delta a_i = \delta b_j
= 0$.

In order to derive the integrable equation for the supermatrix integral, we
now apply the Jacobi identity for a determinant
\begin{equation}
 D \cdot D \left(
	    \begin{array}{cc}
	     i & j \\
	     k & l \\
	    \end{array}
	   \right)
 =
 D \left(
    \begin{array}{c}
     i \\ k
    \end{array}
   \right)
 \cdot
 D \left(
    \begin{array}{c}
     j \\ l
    \end{array}
   \right)
 -
 D \left(
    \begin{array}{c}
     i \\ l
    \end{array}
   \right)
 \cdot
 D \left(
    \begin{array}{c}
     j \\ k
    \end{array}
   \right)
 \, ,
 \label{Jacobi_id}
\end{equation}
where $D$ is a size $n$ determinant, and the size $n-1$ minor determinant
$D \left(\begin{array}{c} i \\ j \end{array} \right)$ is obtained by
removing $i$-th row and $j$-th column from the matrix.
Similarly 
$D \left(\begin{array}{cc} i & j \\ k & l \\ \end{array} \right)$ is the
size $n-2$ determinant obtained by
getting rid of $i, j$-th row and $k,l$-th column.
Setting $i=k=N$ and $j=l=N-1$, we have
\begin{equation}
 \tilde{\cZ}_{N,M} \cdot \tilde{\cZ}_{N-2,M-2}
  =
  \tilde{\cZ}_{N-1,M-1} \cdot \partial_a \partial_b \tilde{\cZ}_{N-1,M-1}
  - \partial_a \tilde{\cZ}_{N-1,M-1} \cdot \partial_b \tilde{\cZ}_{N-1,M-1}
  \, ,
\end{equation}
and by putting $i=k=N-M$ and $j=l=N-M-1$, we obtain
\begin{equation}
 \tilde{\cZ}_{N,M} \cdot \tilde{\cZ}_{N-2,M}
  =
  \tilde{\cZ}_{N-1,M} \cdot \partial^2_a \tilde{\cZ}_{N-1,M}
  - \left( \partial_a \tilde{\cZ}_{N-1,M} \right)^2
  \, .
\end{equation}
These yield the Toda lattice equations~\cite{Hirota:2004}:
\begin{itemize}
 \item 2D Toda equation:
\begin{equation}
  \frac{\tilde{\cZ}_{N+1,M+1} \cdot \tilde{\cZ}_{N-1,M-1}}
      {\left( \tilde{\cZ}_{N,M} \right)^2}
  = 
 \frac{\partial^2}{\partial a \partial b} \log \tilde{\cZ}_{N,M} \, .
\end{equation}
 \item 1D Toda equation:
\begin{equation}
  \frac{\tilde{\cZ}_{N+1,M} \cdot \tilde{\cZ}_{N-1,M}}
      {\left( \tilde{\cZ}_{N,M} \right)^2}
  = 
 \frac{\partial^2}{\partial a^2} \log \tilde{\cZ}_{N,M} \, .
\end{equation}
\end{itemize}
This means that the partition function is the $\tau$-function for the
Toda lattice system both in one and two dimensions simultaneously.
This property can be also found in the ordinary Hermitian matrix
model~\cite{Kimura:2014mua}.

\subsection*{Characteristic polynomial with external source}

We then show that the characteristic polynomial average with the
external source (\ref{SChPoly_det01}) similarly obeys the integrable
equations.
In addition to the external source, we split the parameters for
characteristic polynomials,
\begin{equation}
 \lambda_\alpha = \delta \lambda_\alpha + \lambda \,, \qquad
 \mu_\beta = \delta \mu_\beta + \mu \,, \qquad
 \begin{cases}
  \alpha = 1, \cdots, p \\
  \beta = 1, \cdots, q
 \end{cases}
 \, .
\end{equation}
As well as the previous case, by considering the Taylor expansion
around the center of mass, we obtain the Wronskian-type determinantal
formula in the equal parameter limit $\delta a_i = \delta b_j = \delta
\lambda_\alpha = \delta \mu_\beta = 0$ for $\forall i, j, \alpha, \beta$,
\begin{align}
 \Psi_{N,M;\,p,q} (a,b;\lambda,\mu)
 & = 
 c_{N,M;\,p,q} \,
 \tilde \Psi_{N,M;\,p,q} (a,b;\lambda,\mu)
 \, , \\
 \tilde \Psi_{N,M;\,p,q} (a,b;\lambda,\mu)
 & =
 \det 
 \left(
 \begin{array}{cc}
  Q_{i+k-2}(a) & P_{k-1}^{(\alpha-1)} (\lambda) \\
  R^{(i-1,j-1)}(a;b) & {S}_{\rm L}^{(\alpha-1,j-1)} (\lambda;b) \\
  S_{\rm R}^{(\beta-1,i-1)} (a;\mu) & 
   \tilde{R}^{(\alpha-1,\beta-1)}(\lambda;\mu) \\
 \end{array}
 \right) \, ,
 \label{SChPoly_det02}
\end{align}
with the constant
\begin{equation}
 c_{N,M;\,p,q} = 
  \frac{(a-b)^{NM} (\lambda-\mu)^{pq}}
       {G(N+1) G(M+1) G(p+1) G(q+1)} \, .
\end{equation}
Here $\displaystyle P_{k}^{(\alpha)}(\lambda) =
\frac{d^{\alpha}}{d\lambda^{\alpha}} P_{k}(\lambda)$, 
$\displaystyle S_{\rm L}^{(\alpha,j)}(\lambda;b) =
\frac{\partial^\alpha}{\partial\lambda^\alpha} 
\frac{\partial^j}{\partial b^j}S_{\rm L}(\lambda;b)$, 
and so on.
In this case, since the determinantal formula (\ref{SChPoly_det02})
consists of six blocks, we correspondingly obtain six equations by
applying the identity (\ref{Jacobi_id}):
\begin{align}
 \tilde\Psi_{N,M;\,p,q} \cdot \tilde\Psi_{N,M;\,p-2,q-2}
  & = 
 \tilde\Psi_{N,M;\,p-1,q-1} \cdot 
 \partial_\lambda \partial_\mu\tilde\Psi_{N,M;\,p-1,q-1} 
 - \partial_\lambda \tilde\Psi_{N,M;\,p-1,q-1} \cdot 
 \partial_\mu\tilde\Psi_{N,M;\,p-1,q-1} \, , 
 \label{eq1} \\
 \tilde\Psi_{N,M;\,p,q} \cdot \tilde\Psi_{N-2,M;\,p,q-2}
 & =
 \tilde\Psi_{N-1,M;\,p,q-1} \cdot 
 \partial_a \partial_\mu \tilde\Psi_{N-1,M;\,p,q-1} 
 - \partial_a \tilde\Psi_{N-1,M;\,p,q-1} \cdot 
 \partial_\mu \tilde\Psi_{N-1,M;\,p,q-1} 
 \, , \\
 \tilde\Psi_{N,M;\,p,q} \cdot \tilde\Psi_{N,M-2;\,p-2,q}
 & =
 \tilde\Psi_{N,M-1;\,p-1,q} \cdot 
 \partial_\lambda \partial_b \tilde\Psi_{N,M-1;\,p-1,q} 
 - \partial_\lambda \tilde\Psi_{N,M-1;\,p-1,q} \cdot 
 \partial_b \tilde\Psi_{N,M-1;\,p-1,q} 
 \, , \\
 \tilde\Psi_{N,M;\,p,q} \cdot \tilde\Psi_{N-2,M-2;\,p,q}
 & =
 \tilde\Psi_{N-1,M-1;\,p,q} \cdot 
 \partial_a \partial_b \tilde\Psi_{N-1,M-1;\,p,q} 
 - \partial_a \tilde\Psi_{N-1,M-1;\,p,q} \cdot 
 \partial_b \tilde\Psi_{N-1,M-1;\,p,q} 
 \, , \\
 \tilde\Psi_{N,M;\,p,q} \cdot \tilde\Psi_{N,M;\,p-2,q}
 & =
 (p-1)
 \left(
 \tilde\Psi_{N,M;\,p-1,q} \cdot 
 \partial_a \partial_\lambda \lambda \tilde\Psi_{N,M;\,p-1,q} 
 - \partial_\lambda \tilde\Psi_{N,M;\,p-1,q} \cdot 
 \lambda \partial_a \tilde\Psi_{N,M;\,p-1,q} 
 \right)
 \, , \\
 \tilde\Psi_{N,M;\,p,q} \cdot \tilde\Psi_{N-2,M;\,p,q}
 & =
 p \left(
 \tilde\Psi_{N-1,M;\,p,q} \cdot 
 \lambda \partial^2_a \tilde\Psi_{N,M;\,p-1,q} 
 - \partial_a \tilde\Psi_{N,M;\,p-1,q} \cdot 
 \lambda \partial_a \tilde\Psi_{N,M;\,p-1,q}     
   \right)
 \, .
 \label{eq6}
\end{align}
We have used the following identity to derive some of these
equations 
\begin{eqnarray}
  \det \left(
       \begin{array}{ccc}
	(x^N)^{(0)} & \cdots & (x^{N})^{(M-1)} \\
	\vdots  &  & \vdots \\
	(x^{N+M-2})^{(0)} & \cdots & (x^{N+M-2})^{(M-1)} \\
	(x^{N+M})^{(0)} & \cdots & (x^{N+M})^{(M-1)} \\
       \end{array}
      \right) 
 & = & M \, x \, \det_{1 \le i, j \le M} (x^{N+i-1})^{(j-1)}
 \, ,
 \nonumber \\
\end{eqnarray}
where we denote $\displaystyle (x^j)^{(k)} = \frac{d^k}{dx^k} x^j$.
From the first equation (\ref{eq1}), for example, we obtain the
two-dimensional Toda lattice equation
\begin{equation}
 \frac{\tilde{\Psi}_{N,M;\,p+1,q-1} \cdot \tilde{\Psi}_{N,M;\,p-1,q-1}}
      {\tilde{\Psi}_{N,M;\,p,q}}
 =
 \frac{\partial}{\partial \lambda} \frac{\partial}{\partial \mu} 
 \log \tilde{\Psi}_{N,M;\,p,q}
 \, .
\end{equation}
We can similarly obtain the two-dimensional integrable equations from
others, except for (\ref{eq6}), which leads to the one-dimensional Toda
lattice equation.
Therefore, in this sense, the characteristic polynomial average with the
supermatrix model plays a role of the $\tau$-function for the Toda
lattice system.

Let us comment on higher order integrable equations, which the matrix
partition function would satisfy under generic parametrization of the source.
In order to derive such equations it is convenient to introduce the Miwa
coordinate to the time variables,
\begin{equation}
 t_n(A) = \frac{1}{n} \Tr A^{-n} \, .
\end{equation}
In this sense, since there are two and four kinds of source parameters
in (\ref{SMM03}) and (\ref{SChPoly_det02}), we can correspondingly apply
two and four series of the time variables, respectively.
Although these variables are decoupled in the equal parameter limit, and
we obtain the well-known Toda lattice equations, they might interact with
each other when we consider more generic parametrization.
If so, it would be interesting to study such a situation providing more
involved integrable equations.

\section{Discussion}

In this paper we have studied the supermatrix model involving the
arbitrary potential function $W(x)$ with the external source term.
We have derived the determinantal formula for the partition function and
the characteristic polynomial expectation value with this supermatrix model.
Based on this formula, we have exhibited the duality between the external
source and the characteristic polynomial, and then pointed out that this
duality is just interpreted as Fourier transformation.
We have also shown that the partition function and the characteristic
polynomial average satisfy the Toda lattice equation both in one and two
dimensions, especially in the equal parameter limit.
This implies that we can obtain the corresponding $\tau$-function as
the supermatrix integral with an external source as well as the ordinary
Hermitian matrix model.

As pointed out in Sec.~\ref{sec:ch_poly}, we can obtain various
situations from the supermatrix model by taking the corresponding limit.
From this point of view, it is interesting to study not only simple
reductions of the characteristic polynomial average (\ref{EV_ratio01}),
e.g., $M = 0$, $q=0$, but also the analytic continuation to negative numbers.
Actually the $\U(N|M)$ supermatrix integral is, at least perturbatively,
equivalent to non-supersymmetric matrix model through the analytic
continuation $M \to -M$, but with the two-cut solution, which breaks the
original symmetry, $\U(N+M) \to \U(N) \times \U(M)$.
This kind of equivalence has been recently gathering a great deal of
attention in the area of string/M-theory~\cite{Marino:2009jd}.
For example, it would be interesting to see how the duality relation is
affected by such an analytic continuation.
It is expected that we can obtain more various kinds of correlation
functions, and all the reductions could
preserve the determinantal
structure~\cite{Strahov:2002zu,Bergere:2009zm}, through the analytic
continuation.

From the random matrix theoretical point of view, it is interesting to
extend the present result to other situations, e.g., $\mathrm{OSp}(N|M)$
supermatrix model, chiral ensemble, complex matrix model, two-matrix
model, and so on.
To investigate the matrix model with external source, the
HCIZ formula plays an important role to
integrate out the angular part of the matrix.
Thus it seems better to start with deriving the corresponding formula,
for example, by considering the differential equation for the matrix integral.

\subsection*{Acknowledgements}

We would like to thank B. Eynard for the collaboration on related subjects.
We are also grateful to M. Berg\`ere and E. Br\'ezin for correspondence.
This work is supported in part by Grant-in-Aid for JSPS
Fellows~(\#25-4302).


\bibliographystyle{ytphys}
\bibliography{/Users/k_tar/Dropbox/etc/conf}

\end{document}